\documentclass[twocolumn,preprintnumbers,amsmath,amssymb,superscriptaddress]{revtex4-2}
\UseRawInputEncoding

\usepackage{latexsym,amssymb,amsthm,amsmath,epsfig,braket}
\usepackage[left=2.00cm, right=2.00cm, top=2.00cm, bottom=2.00cm]{geometry}
\usepackage{xcolor}
\usepackage{svg}
\usepackage[colorlinks, citecolor={blue!80!black}, urlcolor={blue!80!black}, linkcolor={red!50!black}]{hyperref}
\usepackage{float}
\usepackage{hyperref}
\hypersetup{
    citecolor=red,
    colorlinks=true,
    linkcolor=blue,
    filecolor=blue,      
    urlcolor=blue,
}

\begin{document}

\title{Photonic spin Hall effect in Haldane model materials}
\author{Muzamil Shah}
\affiliation{Department of Physics, Zhejiang Normal University, Jinhua, Zhejiang 321004, China}
\affiliation{Zhejiang Institute of Photoelectronics \& Zhejiang Institute for Advanced Light Source, Zhejiang Normal University, Jinhua, Zhejiang 321004, China.}
\author{Muhammad Sabieh Anwar}
\affiliation{Department of Physics, Syed Babar Ali School of Science and Engineering, Lahore University of Management Sciences (LUMS), Opposite Sector U, D.H.A., Lahore 54792, Pakistan}
\author{Reza Asgari}
\email{asgari@ipm.ir}
\affiliation{Department of Physics, Zhejiang Normal University, Jinhua, Zhejiang 321004, China}
\affiliation{School of Physics, Institute for Research in Fundamental Sciences (IPM), Tehran 19395-5531, Iran}
\author{Gao Xianlong}
\email{gaoxl@zjnu.edu.cn}
\affiliation{Department of Physics, Zhejiang Normal University, Jinhua, Zhejiang 321004, China}

\begin{abstract}

We investigate the photonic spin Hall effect (PSHE) of light beams reflected from surfaces of various two-dimensional crystalline structures while considering their associated time-reversal $\mathcal{T}$ and inversion $\mathcal{I}$ symmetries. Using the modified Haldane model Hamiltonian with tunable parameters as a generic system, we explore longitudinal and transverse spin-separations of the reflected beam in both topologically non-trivial and trivial systems. The PSHE observed in these materials is attributed to their topology. Topological phase transitions in buckled Xene monolayer materials are demonstrated through the PSHE, showing the manipulation of spin-orbit coupling and external electric fields. Moreover, we investigate spatial shifts in the PSHE of monolayer transition metal dichalcogenides, suggesting that the spin and valley degrees of freedom of charge carriers provide a promising avenue to manipulate the PSHE in both classes of these materials. The study suggests that the PSHE in Haldane model materials can serve as a metrological tool for characterizing topological phase transitions through quantum weak value measurement techniques.

\end{abstract}

\maketitle

\section{ Introduction}
The photonic spin Hall effect (PSHE)~\cite{PhysRevLett.95.136601, leyder2007observation} is the optical counterpart of the spin Hall effect observed in electronic systems. The photon spin (helicity) is affected by a refractive index gradient, while in electronic systems, the analogous phenomena entail the electronic spin responding to an electric potential gradient \cite{onoda2004hall, bliokh2015spin}. The mechanism lying at the heart of the PSHE is the effective spin-orbit coupling (SOC) \cite{dong2020controlling}. In the PSHE, when light with a specific polarization state encounters a surface with characteristic symmetries, the reflected light beams undergo a distinctive split into two parts with opposite spin angular momenta. Although both the PSHE and the Goos-H\"{a}nchen
effect \cite{Goos1947} are optical phenomena, they arise from different mechanisms and have different
forms. The PSHE is a universal phenomenon applicable to any photonic interface, and extensive investigations have been conducted across various physical systems, including optical interfaces \cite{bliokh2015spin}, semiconductors \cite{menard2010ultrafast}, plasmonics \cite{shitrit2011optical}, meta-surfaces \cite{yin2013photonic}, strained Weyl semimetals \cite{jia2021tunable}, topological materials \cite{jia2022spin}, hyperbolic metamaterials \cite{kapitanova2014photonic}, and even in the high-energy physics \cite{gosselin2007spin}. Studying spin-dependent splitting of reflected light provides researchers with valuable insights into the material's electronic properties, offering a unique avenue to probe and comprehend diverse kinds of materials. 

A hexagonal lattice structure, often depicted as an interconnected lattice of hexagons, exhibits symmetry under time-reversal ($\mathcal{T}$) and inversion ($\mathcal{I}$) operations. Pristine graphene, a well-known example with a honeycomb lattice, features two touching bands at distinct points known as Dirac points~\cite{Novoselov2004}. Haldane demonstrated that breaking either the $\mathcal{T}$ or $\mathcal{I}$ symmetry in hexagonal lattices induces a band gap in the electronic structure, transforming the material from the semi-metallic phase to the band insulator phase. What emerges is distinct topological phases in the absence of an external magnetic field~\cite{PhysRevLett.61.2015}. In two-dimensional (2D) hexagonal materials like graphene, buckled Xene monolayers (i.e. silicene, germanene, and stanene),  the preserved symmetries $\mathcal{T}$ and $\mathcal{I}$ give rise to pseudo-spin, while in layered transition metal dichalcogenides (TMDs), and hexagonal boron nitride ($\mathrm{hBN}$) nanosheets, preserved $\mathcal{T}$ symmetry and broken $\mathcal{I}$ symmetry,  give rise to valley quantum degrees of freedom.  Consequently, the interplay of spin and valley dynamics in these diverse range of materials holds substantial promise for applications in valleytronics and spintronics~\cite{schaibley2016valleytronics, pesin2012spintronics, sasaki2008pseudospin}.

Over recent years, the PSHE has been extensively studied in 2D materials. Notably, it has been predicted to be sensitive to the quantized Hall conductivity, allowing its use in probing the quantum Hall effect in monolayer graphene subject to an external magnetic field~\cite{PhysRevA.95.013809,kort2016quantized}.  Furthermore, the PSHE has also been proposed as a tool for probing topological phase transitions in monolayer silicene \cite{PhysRevLett.119.147401,shah2021probing}. Moreover, it has been employed to investigate moiré superlattices and twist angles in 2D systems~\cite{PhysRevB.98.195431,zhang2022photonic,xia2023photonic}. Similarly, enhanced the PSHE on the surface of monolayer black phosphorus in the terahertz region~\cite{lin2018photonic, Zhang:18,JiaLiZhou2020} has been explored. In short, the versatility of the PSHE makes it a valuable tool for exploring the optical properties and topological characteristics of diverse 2D materials~\cite{PhysRevApplied.13.014057, Zhou2012,Chen2017}.

Recently Chen \emph{et al.} successfully detected the PSHE in monolayer $\mathrm{MoS_2}$ and employed weak measurement techniques to extract the optical constants~\cite{chen2021measurement}. Das \emph{et al.} conducted experiments on monolayer $\mathrm{MoS_2}$ positioned on a $\mathrm{Si/SiO_2}$ substrate, revealing that the PSHE is predominantly influenced by the angle of incidence, post-selection angles, and polarization states~\cite{das2022signature}. In a similar vein, Wang \emph{et al.} explored the PSHE in a one-dimensional photonic crystal composed of ultrathin Au films and transition metal dichalcogenides (TMDCs) such as $\mathrm{MoS_2}$, $\mathrm{MoSe_2}$, $\mathrm{WS_2}$, and $\mathrm{WSe_2}$. Their findings demonstrated that the photonic crystal featuring monolayer $\mathrm{WS_2}$ exhibits highly robust PSHE~\cite{das2022signature}. Despite these insightful theoretical and experimental investigations, there currently exists a gap in our knowledge concerning how the PSHE behaves in the presence of spin and valley degrees of freedom exhibited by electrons in TMDCs. Motivated by the distinctive optoelectronic responses of these 2D hexagonal materials, we aim to full this gap within the framework of a general Haldane low-energy model Hamiltonian \cite{PhysRevLett.61.2015,vanderbilt2018berry}. 

We employ a generic Haldane model Hamiltonian within the Kubo formalism to compute optical conductivities. After obtaining the complex functions of the optical conductivities, we proceed to solve Fresnel's reflection coefficients using Maxwell equations. Our developed model provides a comprehensive description of the spin-dependent splitting in reflection on the surface of 2D Haldane model materials. This encompasses both transverse and longitudinal spin-dependent displacements in the PSHE across various 2D hexagonal Haldane model materials. Initially, we meticulously scrutinize the longitudinal and transverse spin-dependent displacements within both topological and trivial systems, providing insights into the sign-switching phenomenon characterizing the PSHE. Subsequently, we broaden our investigation to encompass the realm of buckled Xene monolayer material. In the process, we emphasize the significance of spin-dependent shifts as intricate indicators of topological phase transitions, each unveiling distinct behaviors in various states. Lastly, our exploration delves into the spin and valley-polarized reflected spin-dependent displacements within monolayer transition metal dichalcogenides. Our findings highlight the remarkable sensitivity of the PSHE to spin and valley indices, as well as to the effective mass bands inherent in these materials.

This article is structured as follows: In Sect.~\ref{section:2}, we use the modified Haldane Hamiltonian and derive optical conductivities. In Sect.~\ref{section:3}, we discuss Fresnel's reflection coefficients for an interface coated with 2D Haldane material and the relation of the PSHE. In Sect.~\ref{section:4}, we present and discuss the PSHE in various 2D crystalline materials and finally, we summarize our main results in Sect.~\ref{section:5}. 


\section{The modified low-energy Haldane model Hamiltonian and optical conductivity}\label{section:2}

We consider a generic model Hamiltonian known as the modified Haldane model \cite{PhysRevLett.61.2015,vanderbilt2018berry,pratama2020circular}:
\begin{equation}\label{q1}
\hat{\mathcal{H}} = -t_1 \sum_{\langle i, j\rangle} a_i^{\dagger} a_j + t_2 \sum_{\langle\langle i, j\rangle\rangle} e^{i \nu_{i j} \phi} a_i^{\dagger} a_j + \mathcal{M} \sum_i \chi_i a_i^{\dagger} a_i,
\end{equation}
where, $a_i^{\dagger}$ and $a_i$ are the fermionic creation and annihilation operators of the $i$th atomic site. The tight-binding parameters are $t_1$ and $t_2$. The first term in Eq. \eqref{q1} represents nearest-neighbor $(\mathrm{NN})$ interactions, while the second term captures next-nearest-neighbor (NNN) interactions. This term introduces a breaking of time-reversal symmetry $\mathcal{T}$ for a non-zero phase angle $\phi$ in the exponential factor $e^{i \nu_{i j} \phi}$. Here, $v_{i j} = +1(-1)$ represents the clockwise (anticlockwise) NNN hopping direction. The last term denotes the on-site potential difference for the $A$ and $B$ sublattices, where $\chi_i = +1(-1)$ for the $A(B)$ sublattice. This term is responsible for breaking the $A$ and $B$ sublattice inversion symmetry $\mathcal{I}$, resulting in a band gap of $2 \mathcal{M}$ at the $K$ and $K^{\prime}$ valleys inside the Brillouin zone. It should be noted that our investigations
focus on large systems so that edge effects can be neglected. Taking the Fourier transform of the Hamiltonian, one can obtain the low-energy Hamiltonian near the $K$ and $K'$ points \cite{cooper2019topological,ren2016topological}:
\begin{equation}\label{q2}
\hat{\mathcal{H}}(\boldsymbol{k})=-\lambda+\hbar v_F\left(\tau k_x \hat{\sigma}_x+k_y \hat{\sigma}_y\right)+\Delta_\tau  \hat{\sigma}_{z},
\end{equation}
where, $v_F=\sqrt{3} a t_1 /(2 \hbar)$ is the Fermi velocity, $\tau=+1(-1)$ denotes the $K\left(K^{\prime}\right)$ valley, $\lambda = 3 t_2 \cos \phi$ and $\Delta_\tau = M-\tau 3 \sqrt{3} t_2 \sin \phi$. 
The eigenvalues of the Hamiltonian are given by  $\mathcal{E}^{\tau}_{\eta}(\boldsymbol{k}) = -\lambda + \eta \sqrt{\left(\hbar v_F \boldsymbol{k}\right)^2+\Delta_\tau^2}$ where
$\eta=+1(-1)$ denotes the conduction (valance) band index. The corresponding eigenfunctions can be obtained from Eq. \eqref{q2} as
\begin{equation}\label{q3a}
|\psi_{\eta}^{\tau} (\boldsymbol{k})\rangle=\frac{1}{\sqrt{2\left[\mathcal{E}_{\eta}^{\tau}(\boldsymbol{k})+\lambda\right]}}\left[\begin{array}{c}
\sqrt{\mathcal{E}_{\eta}^{\tau}(\boldsymbol{k})+\lambda+\Delta_\tau} \\
e^{i \varphi / \tau}\sqrt{\mathcal{E}_{\eta}^{\tau}(\boldsymbol{k})+\lambda-\Delta_\tau} 
\end{array}\right],
\end{equation}
where, $\varphi= \mathrm{tan^{-1}} \left(k_y / k_x\right)$.

We calculate the optical conductivity by making use of the Kubo formalism. Therefore, the complex optical conductivity tensor at the $K$ and $K'$ valleys can be obtained as 
\begin{equation}\label{q4}
\sigma_{ij}(\Delta_{\tau},\lambda,\omega)=\sum_{\tau=\pm1}\left(\sigma_{i j}^{\mathrm{intra}}(\Delta_{\tau},\lambda,\omega)+\sigma_{i j}^{\mathrm{inter}}(\Delta_{\tau},\lambda,\omega)\right), 
\end{equation}
where, $i, j$ can be $x$ or $y$. It should be noted that both the interband $\sigma_{ij}^{\mathrm{inter}}$ and intraband $\sigma_{ij}^{\mathrm{intra}}$ conductivities contribute to the total optical response represented by the complex dynamical conductivity $\sigma_{ij}$ of the Haldane model materials. The intraband conductivity  $\sigma_{i j}^{\mathrm{intra}}$ is given by \cite{rodriguez2017casimir,shah2021probing,pratama2020circular}:
\begin{widetext}
\begin{equation}\label{q5}
\begin{aligned}
\sigma_{i j}^{\mathrm{intra}}(\Delta_{\tau},\lambda,\omega)
= & -i e^2 \hbar \int \frac{d^2 \boldsymbol{k}}{(2 \pi)^2}\left(\frac{d f(\mathcal{E}^{\tau}_{v}(\boldsymbol{k}))}{d \mathcal{E}^{\tau}_{v}(\boldsymbol{k}) }\frac{\left\langle \psi_{v}^{\tau}\left|\hat{v}_i\right| \psi_{v}^{\tau}\right\rangle\left\langle \psi_{v}^{\tau}\left|\hat{v}_j\right| \psi_{v}^{\tau}\right\rangle}{ (\omega+i \Gamma)}\right.\left.+\frac{d f(\mathcal{E}^{\tau}_{c}(\boldsymbol{k}))}{d \mathcal{E}^{\tau}_{c}(\boldsymbol{k}) } \frac{\left\langle \psi_{c}^{\tau}\left|\hat{v}_i\right| \psi_{c}^{\tau}\right\rangle\left\langle \psi_{c}^{\tau}\left|\hat{v}_j\right| \psi_{c}^{\tau}\right\rangle}{\hbar (\omega+i \Gamma)}\right),
\end{aligned}
\end{equation}
where $v$  and $c$ denote the valence and conduction bands, respectively, $f(\mathcal{E}^{\tau}_{\eta})$ is the Fermi-Dirac distribution function and  $\hat{v}_i=\hbar^{-1}\partial \hat{\mathcal{H}}/{\hbar \partial k_i}
$ is the velocity operator. Based on symmetry consideration, $\sigma_{x y}^{\mathrm{intra}}(\Delta_{\tau},\lambda,\omega)=-\sigma_{y x}^{\mathrm{intra}}(\Delta_{\tau},\lambda,\omega)=0$. The interband conductivity $\sigma_{i j}^{\mathrm{inter}}$, on the other hand, can be expressed as
\begin{equation}\label{q6}
\begin{aligned}
\sigma_{i j}^{\mathrm{inter}}(\Delta_{\tau},\lambda,\omega)= & -i e^2 \hbar \int \frac{d^2 \boldsymbol{k}}{(2 \pi)^2}\left(\frac{f(\mathcal{E}^{\tau}_{v}(\boldsymbol{k}))-f(\mathcal{E}^{\tau}_{c}(\boldsymbol{k}))}{\mathcal{E}^{\tau}_{v}(\boldsymbol{k})-\mathcal{E}^{\tau}_{c}(\boldsymbol{k})} \frac{\left\langle \psi_{v}^{\tau}\left|\hat{v}_i\right| \psi_{c}^{\tau}\right\rangle\left\langle \psi_{c}^{\tau}\left|\hat{v}_j\right| \psi_{v}^{\tau}\right\rangle}{\mathcal{E}^{\tau}_{v}(\boldsymbol{k})-\mathcal{E}^{\tau}_{c}(\boldsymbol{k})+\hbar (\omega+i \Gamma)}\right. \\
& \left.+\frac{f(\mathcal{E}^{\tau}_{c}(\boldsymbol{k}))-f(\mathcal{E}^{\tau}_{v}(\boldsymbol{k}))}{\mathcal{E}^{\tau}_{c}(\boldsymbol{k})-\mathcal{E}^{\tau}_{v}(\boldsymbol{k})} \frac{\left\langle \psi_{c}^{\tau}\left|\hat{v}_i\right| \psi_{v}^{\tau}\right\rangle\left\langle \psi_{v}^{\tau}\left|\hat{v}_j\right| \psi_{c}^{\tau}\right\rangle}{\mathcal{E}^{\tau}_{c}(\boldsymbol{k})-\mathcal{E}^{\tau}_{v}(\boldsymbol{k})+\hbar (\omega+i \Gamma)}\right).
\end{aligned}
\end{equation}
\end{widetext}
By solving the matrix elements of velocity operators, the longitudinal and transverse optical conductivities at zero-temperature, are calculated for both $K$ and $K'$ valleys and thus we have:
\begin{eqnarray}\label{q7}
\sigma_{x x}^{\mathrm{intra}}(\Delta_{\tau},\lambda,\omega)&=& \frac{e^2}{4 \hbar} \frac{\left(\mu_F+\lambda\right)^2-\Delta_\tau{ }^2}{\left(\mu_F+\lambda\right)}\Big( \delta(\hbar \omega) \nonumber\\
&&+i \frac{1}{ \pi \hbar \omega} \Big) \Theta\left(\mu_F-\mathcal{E}_{\mathrm{co}}^{\tau}\right) \nonumber\\
&=& \sigma_{y y}^{\mathrm{intra}}(\omega)
\end{eqnarray}
\begin{eqnarray}\label{q9}
\sigma_{x x}^{\mathrm{inter}}(\Delta_{\tau},\lambda,\omega)&= & \frac{e^2}{16 \hbar}\left[1+\frac{4 \Delta_\tau^2}{(\hbar \omega)^2}\right]\Big(  \Theta\left[\hbar \omega-2\left(\Lambda_\tau+\lambda\right)\right] \nonumber\\
&&+ \frac{i}{ \pi } \ln \left|\frac{2\left(\Lambda_\tau+\lambda\right)-\hbar \omega}{2\left(\Lambda_\tau+\lambda\right)+\hbar \omega}\right|\Big) \nonumber\\
&& +i \frac{e^2}{4 \pi \hbar^2 \omega} \frac{\Delta_\tau{ }^2}{\left(\Lambda_\tau+\lambda\right)} \nonumber\\
&=& \sigma_{y y}^{\mathrm{inter}}(\Delta_{\tau},\lambda,\omega)
\end{eqnarray}
and finally the optical Hall conductivity along the $xy$ direction is computed as
\begin{eqnarray}\label{q10}
\sigma_{x y}^{\mathrm{inter}}(\Delta_{\tau},\lambda,\omega)&= & \frac{e^2 \tau \Delta_\tau}{4 \pi \hbar^2 \omega} \Big( \ln \left|\frac{2\left(\Lambda_\tau+\lambda\right)-\hbar \omega}{2\left(\Lambda_\tau+\lambda\right)+\hbar \omega}\right|\nonumber\\
&&-i \Theta\left(\hbar \omega-2\left(\Lambda_\tau+\lambda\right)\right)\Big)  \nonumber\\
&= & -\sigma_{y x}^{\mathrm{inter}}(\Delta_{\tau},\lambda,\omega),
\end{eqnarray}
where, $\Theta(x)$ is the Heaviside function, $\Lambda_\tau \equiv \max \left[\mu_F, \mathcal{E}_{\mathrm{co}}^{\tau}\right]$ and $\mathcal{E}_{\mathrm{co}}^{\tau}=-\lambda+\left|\Delta_{\tau} \right|$. When both the time-reversal $\mathcal{T}$ and inversion $\mathcal{I}$ symmetries are conserved i.e., $t_2 = 0, ~\phi = 0$, and $\mathcal{M}= 0$, one recover the optical conductivities of monolayer graphene \cite{PhysRevB.78.085432} indicating the soundness of these computations.

\begin{figure}[t!]
	\centering		
\includegraphics[width=1\linewidth]{ 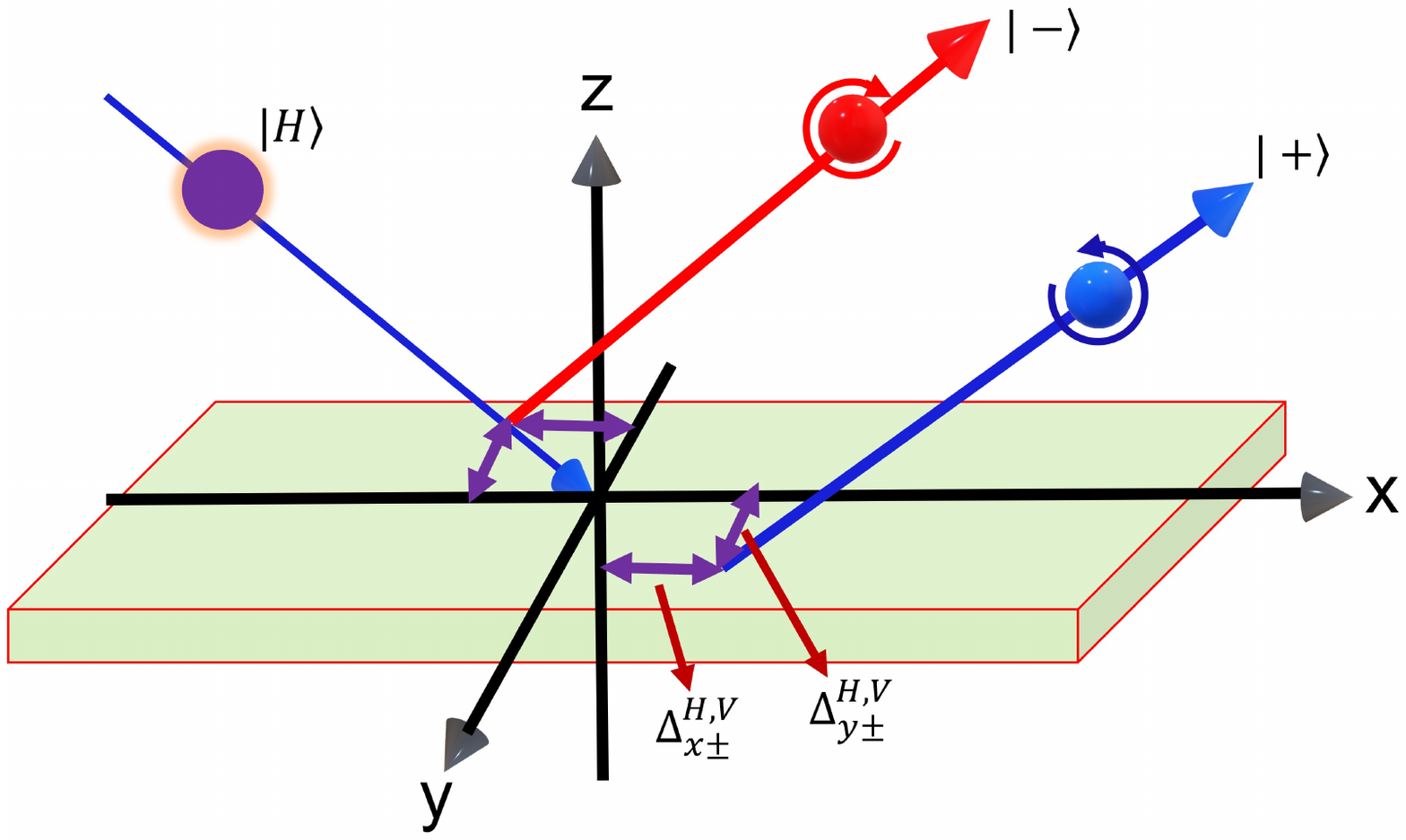}
\caption{Representation of the photonic spin Hall effect of reflected Gaussian light wave from the surface of the Haldane model materials. $\Delta x_{r\pm}^{H/V}$ and $\Delta y_{r\pm}^{H/v}$ are the LHCP/RHCP longitudinal and transverse spin-dependent shifts for the vertical/horizontal polarized optical waves. The horizontal and vertical polarization Gaussian states are denoted by $|H\rangle$ and $|V\rangle$, respectively, while the positive $|+\rangle$ and negative $|-\rangle$ states represent the LHCP and RHCP components.}
\label{PSHE}
\end{figure}

\section{The Photonic spin Hall effect}\label{section:3}
Having calculated the optical conductivities, we can calculate Fresnel's reflection coefficients and the reflected longitudinal and transverse spin-dependent spatial shifts. To do so, we assume that a linearly polarized Gaussian wave of frequency $\omega$ is propagating in air and illuminates the 2D Haldane model material placed on
top of a substrate with an incident angle $\theta_{i}$ as displayed in Fig~\ref{PSHE}. The Cartesian coordinate frame ($x, y, z$) is established, where ($x_i, y_i, z_i$) and ($x_r, y_r, z_r$) present the central wave vectors of the incident and reflected beams, respectively.

To obtain the reflected and transmitted field amplitudes, we apply the appropriate boundary conditions at the surface of the studied system. At the surface, these boundary conditions are given by \cite{kort2015active}:
\begin{eqnarray}\label{q11}
&E_{t}^{s} = E_{i}^{s}+E_{r}^{s}\\
	\label{q12}
&\frac{k_{tz}}{k_{t}}E_{t}^{p}=\frac{k_{iz}}{k_{i}}\big(E_{i}^{p}-E_{r}^{p}\big)\\
	\label{q13}
&\bigg([\frac{k_{tz}}{k_{t}}\sigma_{pp}E_{t}^{p}+\sigma_{ps}E_{t}^{s}\bigg)+\frac{1}{Z_{t}}E_{t}^{p} = \frac{1}{Z_{i}}\big(E_{i}^{p}+E_{r}^{p}\big)\\
	\label{q14}
&\bigg(\frac{k_{tz}}{k_{t}}\sigma_{sp}E_{t}^{p}+\sigma_{ss}E_{t}^{s}\bigg)+\frac{1}{Z_{t}}\frac{k_{tz}}{k_{t}}E_{t}^{s}=\frac{1}{Z_{i}}\frac{k_{iz}}{k_{i}}\big(E_{i}^{p}+E_{r}^{p}\big),
\end{eqnarray}
where, $Z_{0}$ and $Z_{t}=1/\sqrt{\epsilon_{t}}$ are the impedance of the air and the medium. The terms $\sigma_{ss}$, and $\sigma_{ps}$ are the longitudinal and transverse parts of the complex optical conductivities of the system, and $E_{i}^{s}(E_{i}^{p})$, $E_{r}^{s}(E_{r}^{p})$ and $E_{t}^{s}(E_{t}^{p})$ are the complex function of the $s-$ ($p-$) polarized electric fields, respectively. 
Fresnel's reflection coefficients 
of the system
are
\cite{shah2021probing,shah2022electrically,wu2020weak}:
\begin{eqnarray}\label{q15}
	r_{pp}(k,i\omega) &=& \frac{\alpha_{-}\beta_{+}+\sigma_{ps}\sigma_{sp}}{\alpha_{+}\beta_{+}+\sigma_{ps}\sigma_{sp}}, \\
	\label{q16}
	r_{ss}(k,i\omega) &=& \frac{\alpha_{+}\beta_{-}-\sigma_{ps}\sigma_{sp}}{\alpha_{+}\beta_{+}+\sigma_{ps}\sigma_{sp}}, 
	\\
	\label{q17}
	r_{ps}(k,i\omega)&=&-r_{sp}(k,i\omega)=\frac{2}{Z_{i}}\frac{\sigma_{sp}}{\alpha_{+}\beta_{+}+\sigma_{ps}\sigma_{sp}}
\end{eqnarray}
where
\begin{eqnarray}\label{q18}
	\alpha_{\pm} &=&\frac{\sigma_{pp}k_{iz}}{k_{i}}+\frac{k_{iz}k_{t}}{Z_{t}k_{i}k_{tz}}\pm \frac{1}{Z_{i}}, \\
	\label{da27}
	\beta_{\pm} &=& \mp\frac{\sigma_{ss}k_{i}}{k_{iz}}+\frac{k_{i}k_{t}}{Z_{t}k_{iz}k_{t}}-\frac{1}{Z_{i}} \cdot
\end{eqnarray}
The angular spectrum of the incident Gaussian wave in the momentum space with beam waist $w_0$ can be expressed as
$\tilde{E}_i=\frac{w_0} {\sqrt{2 \pi}}\exp \left[\frac{-w_0^2\left(\kappa_{i x}^2+\kappa_{i y}^2\right) }{ 4}\right]$
where the wave vectors of the incident light are denoted by $\kappa_{i x}$ and $\kappa_{i y}$, respectively \cite{wu2020weak}. The horizontal and vertical incident field components in the spin basis can be expressed as
$\tilde{E}_i^H={\left(\tilde{E}_{i+}+\tilde{E}_{i_{-}}\right) }/{\sqrt{2}}$
and
$\tilde{E}_i^V={i\left(\tilde{E}_{i_{-}}-\tilde{E}_{i+}\right)}/{\sqrt{2}}$. 
The amplitudes of the reflected angular spectra with the horizontal ($H$) and vertical ($V$) polarization Gaussian states are linked to those of the incident wave by the transfer matrix as follows  \cite{shah2021probing,jia2021tunable}:
\begin{equation}\label{q22}
\begin{bmatrix} {\tilde{E}_r^{H}}
		\\  {\tilde{E}_r^{V}} \\ \end{bmatrix}=\begin{bmatrix} r_{pp} -\frac{\kappa_{ry}~\rho~\cot(\theta_{in})}{k_{0}}  & r_{ps}+\frac{\kappa_{ry}~\varphi~\cot(\theta_{in})}{k_{0}} \\ r_{ps}+\frac{\kappa_{ry}~\varphi~\cot(\theta_{in})}{k_{0}} & r_{ss}-\frac{\kappa_{ry}~ \rho~\cot(\theta_{in})}{k_{0}} \end{bmatrix} \begin{bmatrix} {\tilde{E}_i^{H}}
		\\  {\tilde{E}_i^{V}} \\ \end{bmatrix},
\end{equation}
where,  $\varphi=(r_{pp}+r_{ss})$, $\rho=(r_{ps}-r_{sp})$ and $k_0=\omega \sqrt{\varepsilon_0} / c$. Here, for the angular spectrum, the boundary conditions  $\kappa_{rx}=-\kappa_{ix}$ and $\kappa_{ry}=\kappa_{iy}$ are introduced.
By transforming the horizontal and vertical electric field amplitudes of light beams
from the wave vector space into the coordinate space, one may obtain the reflected longitudinal
and transverse spatial shifts as
\cite{wu2020weak,shah2021probing,jia2021tunable},
\begin{eqnarray}\label{q23}
\Delta x_{r\pm}^{H}&=&\mp\frac{1}{k_{0}}\textrm{Re}\bigg(\frac{r_{pp}}{\Psi}\frac{\partial r_{sp}}{\partial \theta_{in}}
	-\frac{r_{sp}}{\Psi}\frac{\partial r_{pp}}{\partial \theta_{in}}\bigg),\\
	\label{q24}
\Delta x_{r\pm}^{V}&=&\mp\frac{1}{k_{0}}\textrm{Re}\bigg(\frac{r_{ps}}{\Psi}\frac{\partial r_{ss}}{\partial \theta_{in}}
	-\frac{r_{ss}}{\Psi}\frac{\partial r_{ps}}{\partial \theta_{in}}\bigg),\\
\Delta y_{r\pm}^{H}&=&\mp\frac{\cot \theta_{i}}{k_{0}}\textrm{Re}\bigg(\frac{\varphi ~r_{pp}}{\Psi}-\frac{\rho ~r_{sp}}{\Psi}\bigg),\\
	\label{q25}
\Delta y_{r\pm}^{V}&=&\mp\frac{\cot \theta_{i}}{k_{0}}\textrm{Re}\bigg(\frac{\varphi ~r_{ss}}{\Psi}+\frac{\rho~r_{ps}}{\Psi}\bigg),
\end{eqnarray}
where, $\Psi=(r_{pp}^2+r_{sp}^2)$. In the
following, for the sake of simplicity, we restrict our discussion to the $H$ polarized incident light wave only for the calculation of the PSHE.
\\
\section{The photonic spin Hall effect in various 2D materials}\label{section:4}
\subsection{Topological and trivial insulator}
\begin{table*}[t]
	{\caption{Optical transitions in different Haldane materials in the $K$ valley. The $\times$ sign represents no transition. The spin-orbit coupling in silicene and $\mathrm{MoS_{2}}$ are $\Delta_{so}$= 3.9 meV and $\Delta_{\mathrm{TMD}}$= 75 meV, respectively. \label{mytab} }}
	\centering
	\begin{tabular}{lllllll}
		\hline 
		&~~Haldane material &~~~ $t_{2}$&~~~~~~~$\mathcal{M}$&~~~~$\phi$&~~~~$T_{1}$&~~~~$T_{2}$      \\ \hline
 		&~~~~~~Topological &~~~~~~0.05 eV &~~~~ 0 eV  &~~~~$\pi/2$ &~~~~$6\sqrt{3}t_{2}$ &~~~~$\times$     \\
		&~~~~~~Trivial &~~~~~~0.05 eV &~~~~ 0.5 eV  &~~~~$0$ &~~~~$2\mathcal{M}$ &~~~~$\times$      \\
		&~~~~~~Buckled Xene (QSHI phase)&~~~~~~$\Delta_{so}/3\sqrt{3}$ meV &~~~~ $\Delta_{so}/2$ meV  &~~~~$\pi/2(-\pi/2)$ &~~~~$\Delta_{so}(\uparrow)$ &~~~~$3\Delta_{so}(\downarrow)$          \\
		&~~~~~~Buckled Xene (VSPM phase) &~~~~~~$\Delta_{so}/3\sqrt{3}$ meV &~~~~ $\Delta_{so}$ meV  &~~~~$\pi/2(-\pi/2)$ &~~~~$\times(\uparrow)$ &~~~~$4\Delta_{so}(\downarrow)$   \\
		&~~~~~~Buckled Xene (BI phase) &~~~~~~$\Delta_{so}/3\sqrt{3}$ meV &~~~~ $2\Delta_{so}$ meV  &~~~~$\pi/2(-\pi/2)$ &~~~~$2\Delta_{so}(\uparrow)$ &~~~~$6\Delta_{so}(\downarrow)$    \\
		&~~~~~~TMDC ($\mathrm{MoS_{2}}$) &~~~~~~$\Delta_{\mathrm{TMD}}/3\sqrt{3}$ eV &~~~~ $\Delta/2$ eV  &~~~~$5\pi/6(-\pi/6)$ &~~~~$(\Delta- \Delta_{\mathrm{TMD}})(\uparrow)$ &~~~~$(\Delta+ \Delta_{\mathrm{TMD}})(\downarrow)$    \\ \hline
	\end{tabular}
\end{table*}
\begin{figure}[]
	\centering		
\includegraphics[width=0.850\linewidth]{ 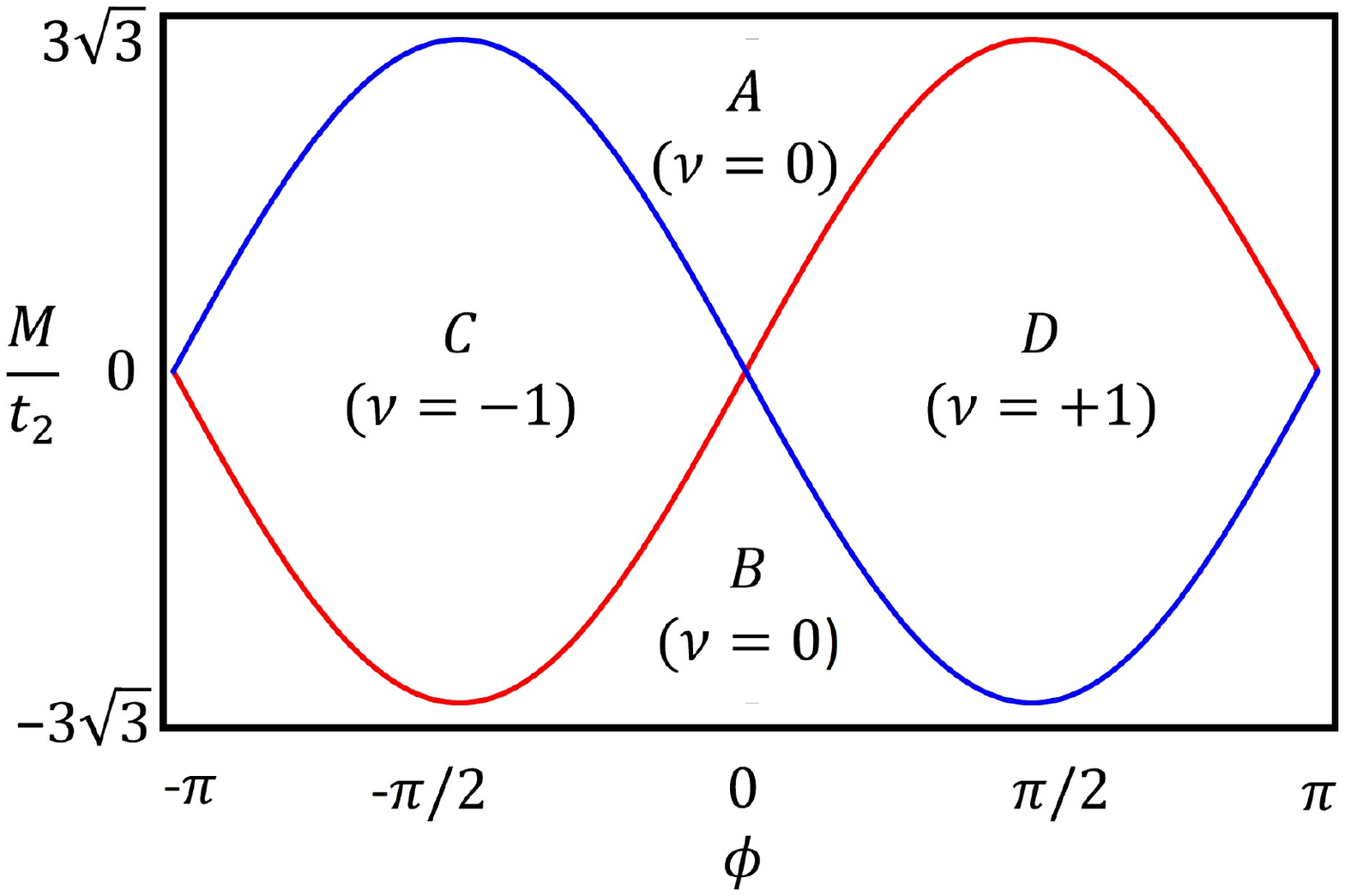}
\caption{Phase
diagram of the general model. The normal and (topological) insulator regime is denoted by $\mathcal{A}$ and $\mathcal{B}$ ($\mathcal{C}$ and $\mathcal{D}$). The blue (red) curves correspond to $\mathcal{M}=\tau 3 \sqrt{3} t_2 \sin \phi$ ($\mathcal{M}=-\tau 3 \sqrt{3} t_2 \sin \phi$), where both the valleys $K$ and $K'$ have zero energy gap. The topological and normal electronic phases and the associated Chern numbers are also indicated.}
\label{phase}
\end{figure}

First, we discuss the phase diagram expressed in $\mathcal{\phi}$ and $\mathcal{M}/t_{2}$ plane as shown in Fig.~\ref{phase}. The blue (red) curve indicates the specific case that the $K\left(K^{\prime}\right)$ valley does not have an energy gap while the $K^{\prime}(K)$ valley has an energy gap. It must be noted that the topological phase transition occurs when the sign of $\mathcal{M}/t_{2}$ changes. When $\left|\mathcal{M} / t_2\right|$ is smaller than $3\sqrt{3}|\sin\phi|$, the phase of the material is topological. On the other hand, if $\left|\mathcal{M} / t_2\right|$ is greater than $3\sqrt{3}|\sin \phi|$, then it is a band insulator (trivial). We specify the normal and (topological) insulator phase with the associated Chern number in regions $\mathcal{A}$ and $\mathcal{B}$ ($\mathcal{C}$ and $\mathcal{D}$) as shown in the phase diagram (see Fig.~\ref{phase}). Although the phase diagram for the Haldane model is well-studied, it is worth mentioning that the topological phases of all the materials examined here are determined by essential parameters such as $t_2$ and $\phi$. Consequently, the topological phases also vary with these parameters. The PSHE observed in these materials is attributed to topology, highlighting the manipulation through spin-orbit coupling and external electric fields. The information on optical conductivities of 2D Haldane model materials is inherited in their energy dispersions. The parameters for the topological and trivial cases are also given in Table \ref{mytab}.
\begin{figure}[]
	\centering		
	\includegraphics[width=0.4850\linewidth]{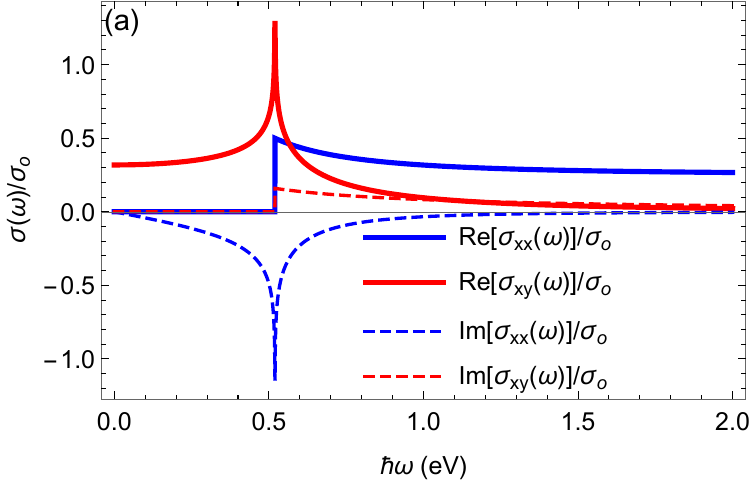}
	\includegraphics[width=0.480\linewidth]{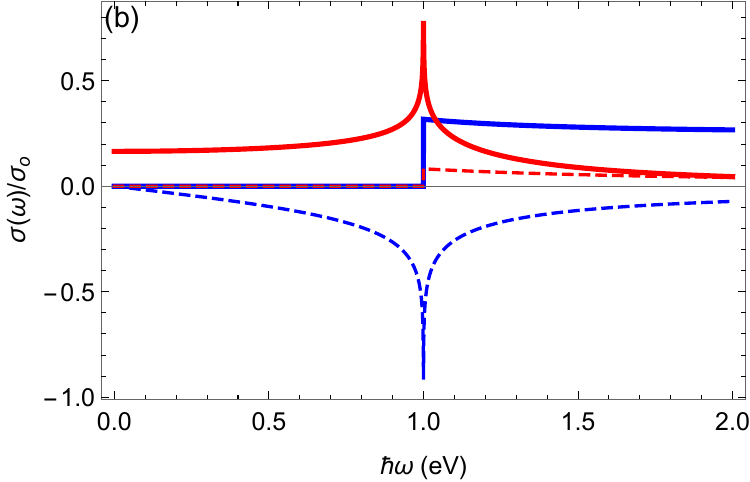}\\
	\includegraphics[width=0.4850\linewidth]{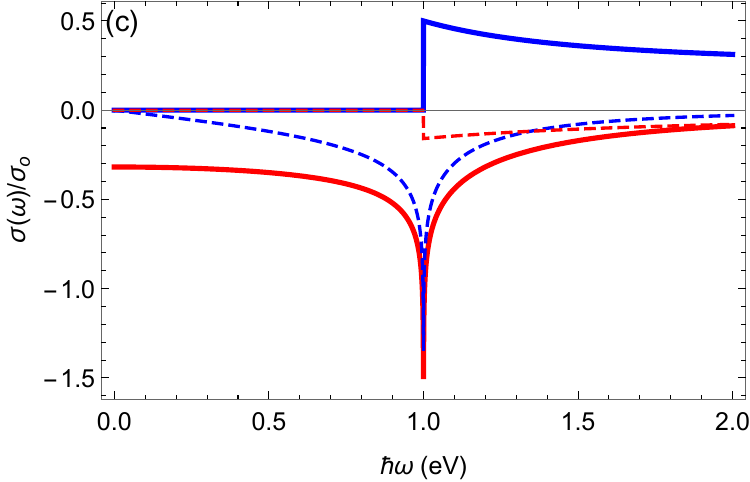}
	\includegraphics[width=0.4850\linewidth]{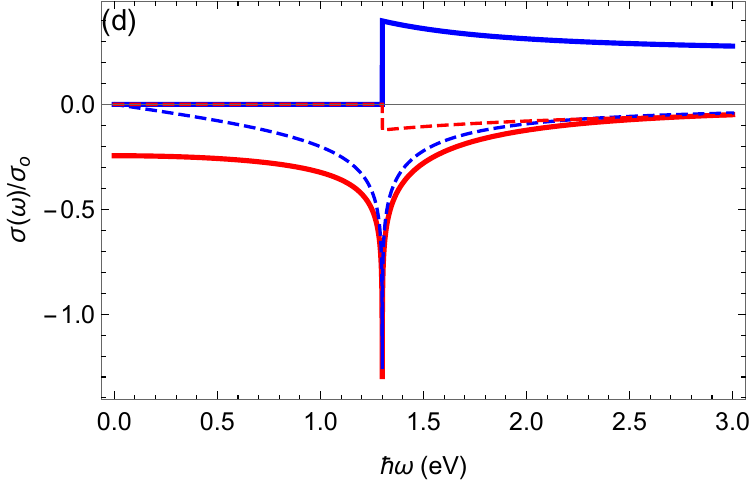}
\caption{The real and imaginary parts of  $\sigma/\sigma_{0}$ for topological case (a) and (b) are given for $\mu_F=0$ while (c) and (d))) are given for $\mu_F=0.5$ eV, respectively. We set a topological ($t_{2}=50$ meV, $\mathcal{M}=0$ and $\mathcal{\phi}=\pi/2$) and (b) a trivial ($t_{2}=50$ meV, $\mathcal{M}=0.5$ eV and $\mathcal{\phi}=0$) cases. The topological cases are shown in (a) and (b) however, the trivial cases are shown in (c) and (d).}
\label{cond1}
\end{figure}
\begin{figure}[]
	\centering		
	\includegraphics[width=0.75\linewidth]{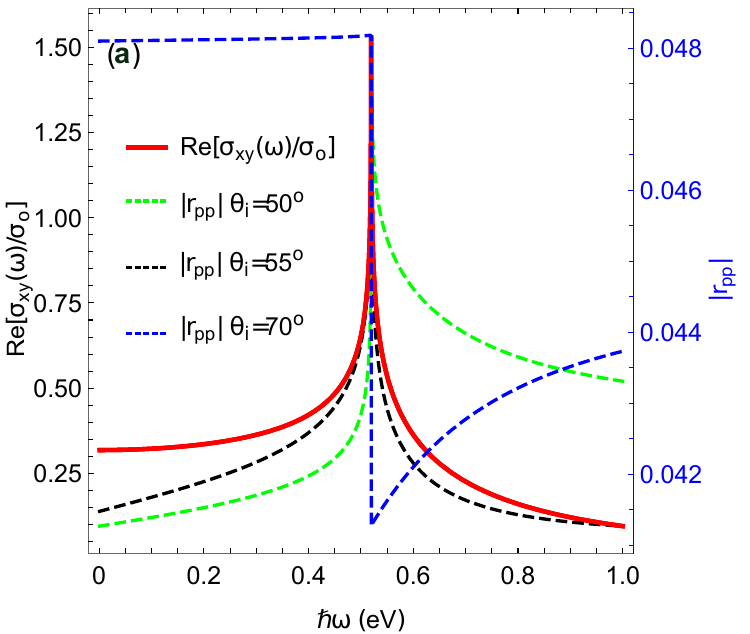}
	\\ \includegraphics[width=0.75\linewidth]{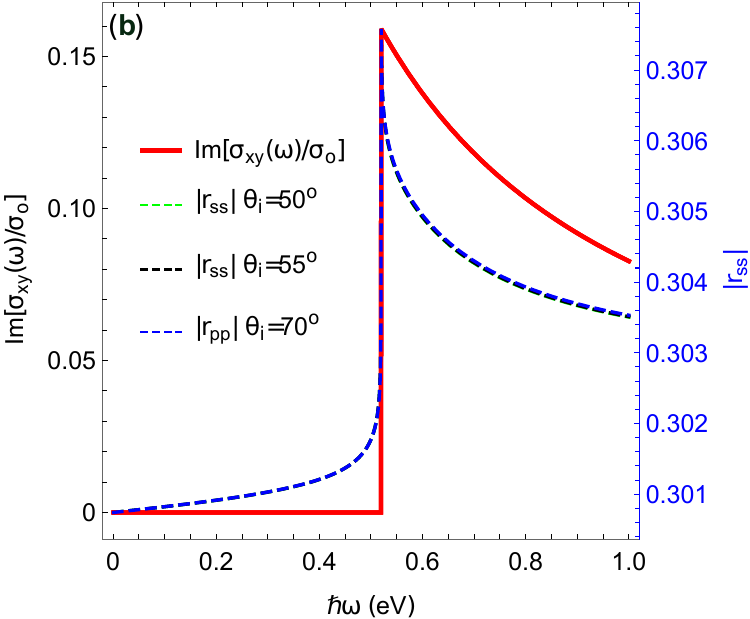}
	\\ \includegraphics[width=0.75\linewidth]{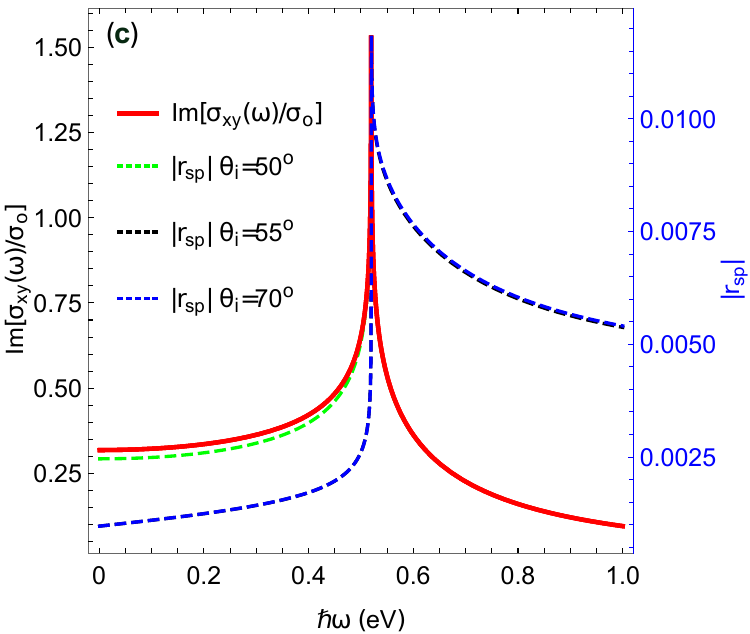}
\caption{Comparisons between Fresnel reflection coefficients $|r_{pp}|$, $|r_{ss}|$, and $|r_{sp}|$ and the optical conductivities
$\sigma_{xy}$ and $\sigma_{xx}$. The $|r_{pp}|$, $|r_{ss}|$ and $|r_{sp}|$ spectra in (a), (b), and (c) are calculated at three different angles of incidence. }
\label{fig2}
\end{figure}

\begin{figure}[t]
	\centering		
\includegraphics[width=0.485\linewidth]{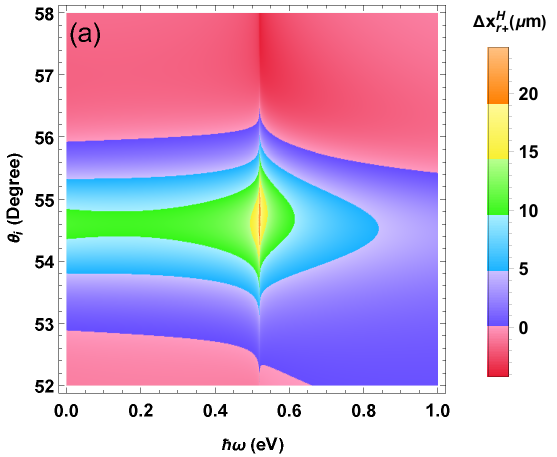}
	\includegraphics[width=0.485\linewidth]{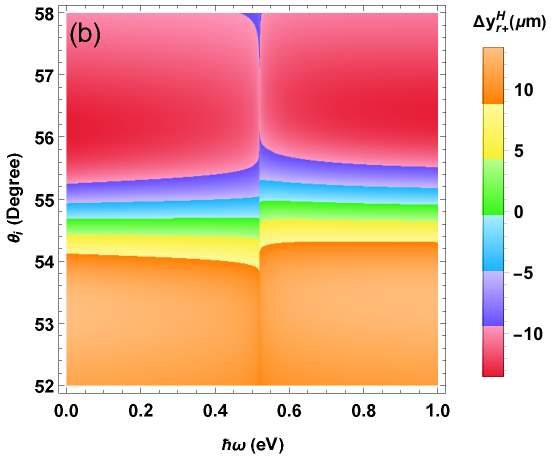}\\
	\includegraphics[width=0.485\linewidth]{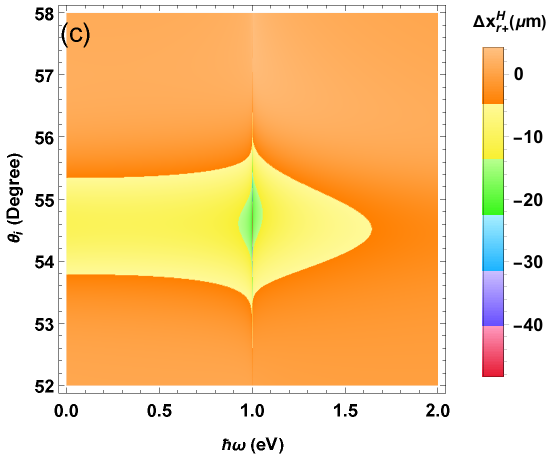}
	\includegraphics[width=0.485\linewidth]{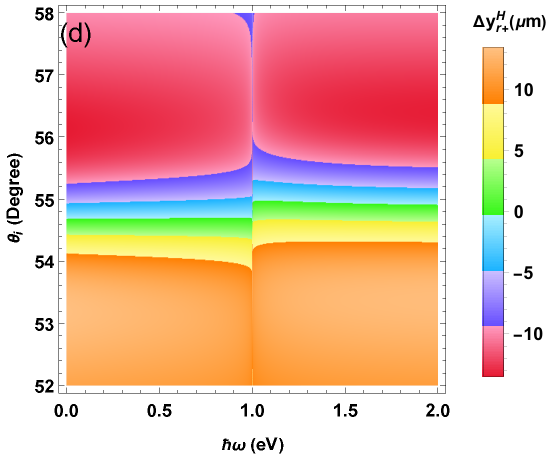}
\caption{Color maps of (a) $\Delta x_{r+}^{H}$ and  (b) $\Delta y_{r+}^{H}$ versus photonic energy and angle of incidence in the $K$ valley in the topological case. Color maps of the (c) $\Delta x_{r+}^{H}$ and  (d) $\Delta y_{r+}^{H}$ versus $\hbar\omega$ and $\theta_{i}$ in the $K$ valley in the trivial case.}
\label{fig22}
\end{figure}

By examining the energy dispersion of the topological and trivial cases, we can delve into the discussion of optical conductivities, the properties of which are rooted in the specified regimes of the phase diagram. In Figs. \ref{cond1}(a)-(d), we depict the optical conductivities for the topological and trivial cases as a function of photon energy in the $K$ valley. In Fig.~\ref{cond1}(a), the real and imaginary components of $\sigma_{xx}$ and $\sigma_{xy}$ are plotted against $\hbar\omega$ for the topological case. The parameter values are detailed in Table \ref{mytab}. We consider the undoped case, where the Fermi energy $\mu_{F}=0$, resulting in pure interband optical transitions. Logarithmic singularities in $\sigma_{xx}$ and $\sigma_{xy}$ conductivities are evident at $\mathcal{E}_{g}=\hbar\omega$, occurring when the incident photon energy, $\hbar\omega$, matches $\mathcal{E}_{g}$. Additionally, the real component of $\sigma_{xx}$ and the imaginary part of $\sigma_{xy}$ monotonically decrease for photon energies higher than $\mathcal{E}_{g}$. However, logarithmic singularities are observed in the imaginary part of $\sigma_{xx}$ and the real component of $\sigma_{xx}$ due to interband transitions. It is noteworthy that for a finite electron doping $\mu_{F}$=0.5 eV, the peaks in optical conductivities are blue-shifted, as illustrated in Fig.~\ref{cond1}(b). For the trivial case, we set $t_2=0.05~\mathrm{eV}$, $\phi=0$, and $\mathcal{M}=0.5~\mathrm{eV}$, corresponding to the $\mathcal{A}$ and $\mathcal{B}$ regions. Figures \ref{cond1}(c) and (d) present the real and imaginary components of the optical conductivities concerning photon energy for the $K$ valley, considering $\mu_{F}$=0 eV and $\mu_{F}$=0.5 eV, respectively. Due to the finite value of the band gap $\mathcal{M}$, the excitation energies of intraband transitions are shifted to higher photon energies.
As a first step, we focus our attention on investigating the dependence of Fresnel reflection coefficients on the complex optical conductivity of Haldane materials.
Our numerical results show that $|r_{pp}|$ attains a minimum value at a specific angle of incidence referred to as the pseudo-Brewster angle $\theta_{B}$ and then rises to the maximum value. In Fig.~\ref{fig2}(a), we provide comparisons between $|r_{pp}|$ and the real part of $\sigma_{xy}$ for three different angles of incidence. We observe that $|r_{pp}|$ and the real part of $\sigma_{xy}$ exhibit a similar trend with $\hbar\omega$ for $\theta_{i} \le \theta_{B}$, but they show different behavior for $\theta_{i} \ge \theta_{B}$, as shown in Fig.~\ref{fig2}(a).  On the other hand, $|r_{ss}|$ and the imaginary component of $\sigma_{xy}$ (or real part of $\sigma_{xx}$) display a similar tendency for three different angles of incidence as shown in Fig.~\ref{fig2}(b). The step-like nature of $|r_{ss}|$ is inherited from the imaginary component of $\sigma_{xy}$ (or real part of $\sigma_{xx}$).  Similarly, $|r_{sp}|$ and the imaginary part of $\sigma_{xy}$ exhibit the a similar pattern for $\theta_{i} \le \theta_{B}$ and in the vicinity of $\theta_{B}$, while at a larger angle of incidence ($\theta_{i}=70^\circ$), a slightly different pattern can be observed, as depicted in Fig.~\ref{fig2}(c). From this visualization, it is evident that the reflection coefficients are highly sensitive to changes in the optical conductivity of Haldane materials.

In the next step, we delve into the longitudinal and transverse spin-splitting displacements of Haldane model materials for both topological and trivial cases. Previous studies \cite{jia2021tunable} have highlighted the sensitivity of these displacements to the minimum values of $|r_{pp}|$ with significant splitting occurring around $\theta_{B}$. To illustrate this, color maps of $\Delta x_{r+}^{H}$ and $\Delta y_{r+}^{H}$ versus $\theta_{i}$ and $\hbar\omega$ are presented in Figs.~\ref{fig22}(a) and (b) for the topological case. It must be noted that for all cases, the in-plane and transverse displacements give extreme values around $\theta_{B}$=$54.6^\circ$. For this purpose, the incident angles for $\Delta x_{r+}^{H}$ and $\Delta y_{r+}^{H}$ in rest of the figures are restricted to 52--$58^\circ$ range. For better clarity,  the color map of LHCP longitudinal PSHE displays extreme displacement at $\hbar\omega=0.5$ eV in the vicinity of $\theta_{B}$, where $|r_{pp}|$ exhibits local minima as shown in Fig.~\ref{fig22}(a). The positive LHCP longitudinal spin-dependent shift $\Delta x_{r+}^{H}$ aligns with the positive peak in the real part of $\sigma_{xy}$ and the Fresnel reflection coefficient $r_{pp}$ near $\theta_{B}$, compare with Fig.~\ref{fig2}(a). Moving to Fig.~\ref{fig22}(b), the transverse displacement $\Delta y_{r+}^{H}$ for the topological case is presented. The step-like behavior of the reflected spin-dependent transverse shift is attributed to $|r_{ss}|$, exhibiting a similar trend with angles of incidence (see Fig.~\ref{fig2}(b)). Notably, $\Delta y_{r+}^{H}$ is positive when the angle of incidence is less than $\theta_{B}$ and becomes negative otherwise. Similarly, Figs. \ref{fig22}(c) and (d) present the color maps of $\Delta x_{r+}^{H}$ and $\Delta y_{r+}^{H}$ for the trivial case, respectively. Both displacements exhibit similar characteristics, with $\Delta x_{r+}^{H}$ being negative due to the negative real part of $\sigma_{xy}$. In the vicinity of $\theta_{B}$, $\Delta x_{r+}^{H}$ displays large values at $\hbar\omega=1$ eV, as shown in Fig.~\ref{fig2}(a). These shifts in both topological and trivial cases are incident energy dependent, illustrating the energy-dependent optical longitudinal and Hall conductivities of Haldane materials.
\\
\subsection{Buckled Xene monolayer materials}
Buckled Xene ($\mathrm{X}$ refers to group-IV elements) 
monolayer materials are silicon (silicene), germanium (germanene), tin (stanene), and lead (plumbene) analogs of
graphene. The major difference is that these materials have out-of-plane structures leading to a large spin-orbit coupling (SOC). The strength of the SOC $\Delta_{so}$ in  silicene \cite{ezawa2012spin}, germanene \cite{liu2011quantum}  and tinene \cite{xu2013large} are 1.55--7.9
meV, 24--93 meV, and 100 meV, respectively. An external electric field $E_{z}$
vertical to the buckled Xene plane breaks the inversion symmetry and results in a staggered potential $\Delta_{z}=elE_z$
between sublattices $A$ and $B$. 
Subsequently, the interaction of the $E_{z}$ with buckled Xene monolayer materials makes the surface Dirac fermions mass controllable at both valleys, which leads to multiple topological phase transitions. If we substitute $\mathcal{M}=\Delta_{z}$, $t_2 = \Delta_{so}/3\sqrt{3}$, and $\phi=\pm\pi/2$, in the dispersion relation of the generic Hamiltonian, we can reproduce the dispersion relation of the buckled Xene monolayer materials as \cite{ezawa2012spin}
$\mathcal{E}^{\tau, s}_{\eta}(\boldsymbol{k})=
 \eta \sqrt{\left(\hbar v_F \boldsymbol{k}\right)^2+(\Delta_{\tau}^{s})^2}$
where, $\Delta_{\tau}^{s}=\Delta_{z}-\tau s \Delta_{so}$ is the energy gap in the buckled Xene monolayer materials, which can
be tuned by variation in $E_z$. The terms $\phi=\pi/2$ and $\phi=-\pi/2$ denote real spin degrees of freedom for spin-up and spin-down electrons, respectively. 

The electronic band structures of materials in the graphene family can be understood by the aforementioned equation for the spin-up and spin-down in the $K$ valley. Notice that spin-dependent electron significantly influences the band structure of buckled Xene and transition metal dichalcogenides materials by generating pseudo-spin polarized bands and spin-
polarized valleys near the Dirac points. Subsequently, the PSHE depends on the electron spin degree of freedom. For the case $\Delta_{z}=0$, the spin-up and spin-down energy bands are degenerate and separated by a gap of $2\Delta_{so}$. For the $\Delta_{z}=0.5 \Delta_{so}$ case, the buckled Xene monolayer material behaves as a quantum spin Hall insulator (QSHI) \cite{ezawa2012spin}. For $\Delta_{z}=\Delta_{so}$, the spin-up electron band gap closes and the buckled Xene monolayer material (which is silicene in this case) undergoes a
topological phase transition from the QSHI phase to the valley-spin polarized metal (VSPM) phase.  
Further increasing the strength of the staggered potential (i.e., $\Delta_{z}>\Delta_{so}$) results in reopening the spin-up and down gaps and the material reaches the band insulator (BI) phase. As a result,
the buckled Xene monolayer materials present n extreme rich phase diagram.

\begin{figure}[]
	\centering		
 \includegraphics[width=0.4850\linewidth]{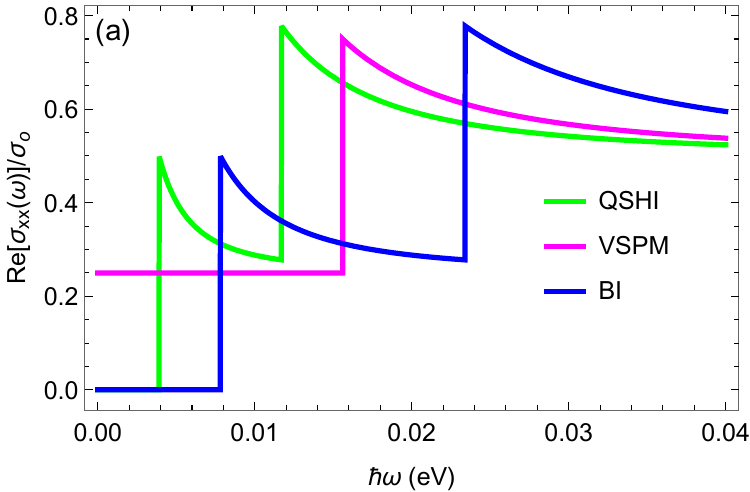}
	\includegraphics[width=0.4850\linewidth]{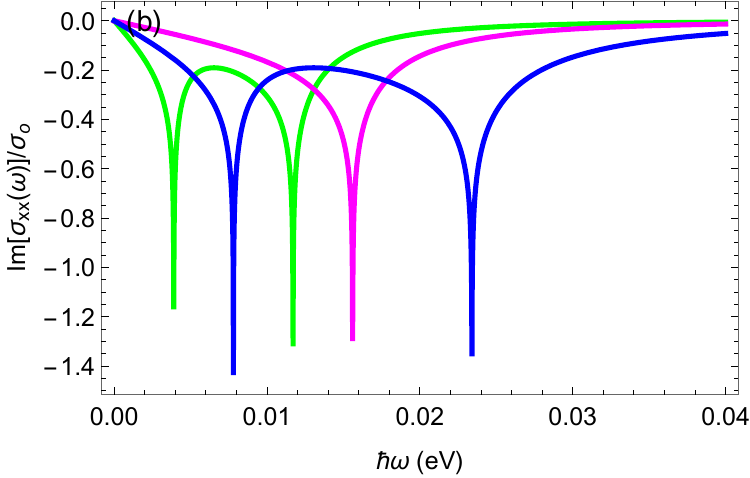}\\
	\includegraphics[width=0.4850\linewidth]{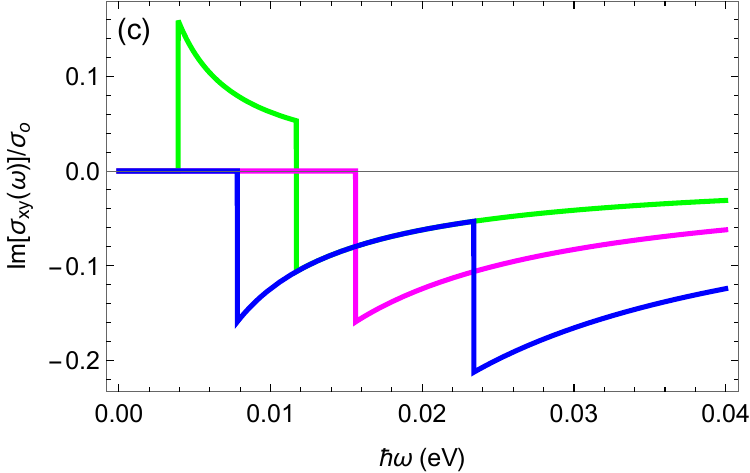}
	\includegraphics[width=0.4850\linewidth]{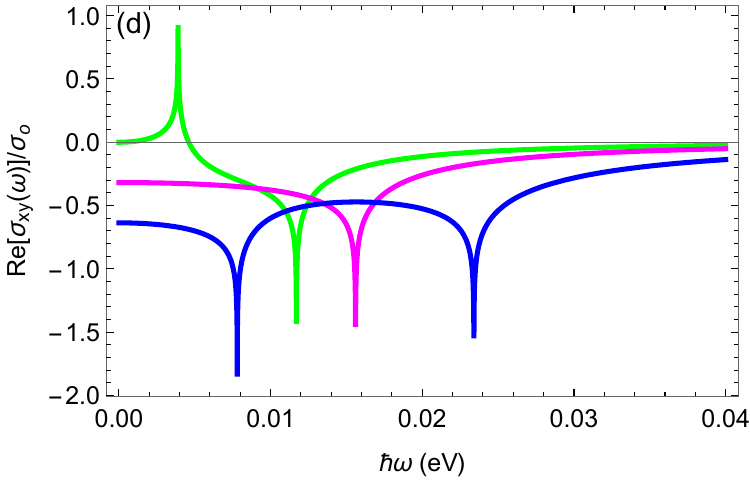}
	\caption{ The real and imaginary components of the optical conductivities versus photon energy for the spin-up and spin-down electrons in the $K$ valley in distinct topological phases. (a) Real parts of $\sigma_{xx}(\omega)/ \sigma_{o}$. (b) Imaginary components of $\sigma_{xx}(\omega)/ \sigma_{o}$. (c) Real parts of $\sigma_{xy}(\omega)/ \sigma_{o}$. (d) Imaginary components of $\sigma_{xy}(\omega)/ \sigma_{o}$. For these simulations, we used $t_2 = \Delta_{so}/3\sqrt{3}$, $\Delta_{so}=3.9$ meV,  $\mathcal{M}=\Delta_{z}$ and $\phi=\pm\pi/2$ for spin-up(spin-down).}
	\label{ENERGY}
\end{figure}

In Figs.~\ref{ENERGY}(a)-(d), ${\sigma}_{xx}$  and ${\sigma}_{xy}$ for spin-up and spin-down electrons in the $K$ valley in the aforementioned topological phases are shown.  In the QSHI phase, there are two steps (jumps) in the longitudinal conductivity at two different photon energies ($\hbar\omega=\Delta_{so}$ and $\hbar\omega=3\Delta_{so}$) as shown in Fig.~\ref{ENERGY}(a). The optical transitions in distinct phases are shown in Table \ref{mytab}. When $\Delta_{z}=\Delta_{so}$, the gap of one of the spin-split bands closes. In the VSPM phase, we can see only one singular peak or step of ${\sigma}_{xx}$  and ${\sigma}_{xy}$ which originates from the interband transitions of the spin-down electron in the $K$ valley at photon energy $\hbar\omega=4\Delta_{so}$. It must be noted that the VSPM instance divides into two topological phases. In the BI phase, two resonant peaks emerge in the conductivity spectra at $\hbar\omega=\Delta_{so}$ and $\hbar\omega=6\Delta_{so}$. 
\begin{figure}[h]
	\centering		
\includegraphics[width=0.485\linewidth]{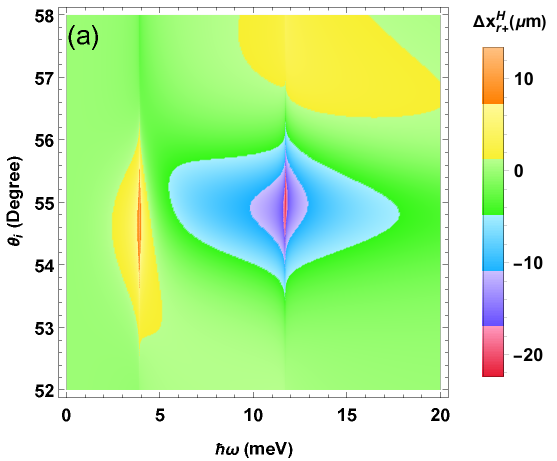}
\includegraphics[width=0.485\linewidth]{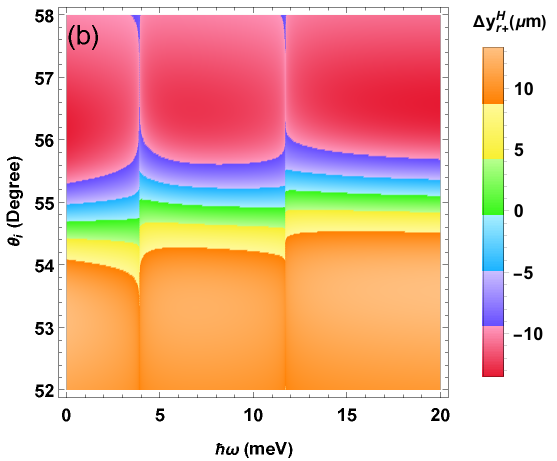}\\
\includegraphics[width=0.485\linewidth]{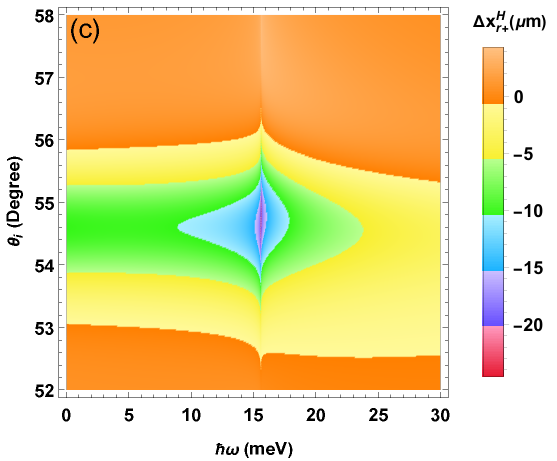}
\includegraphics[width=0.485\linewidth]{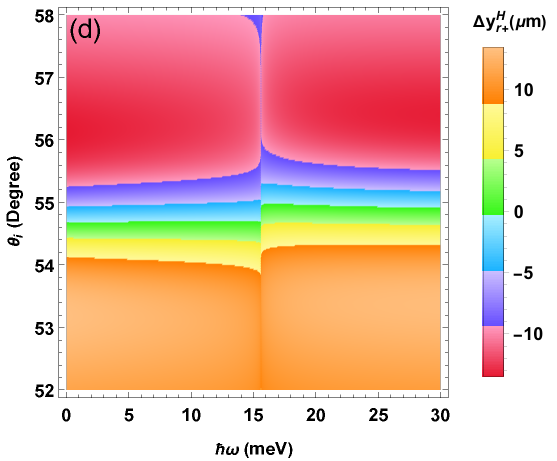}\\
	\includegraphics[width=0.485\linewidth]{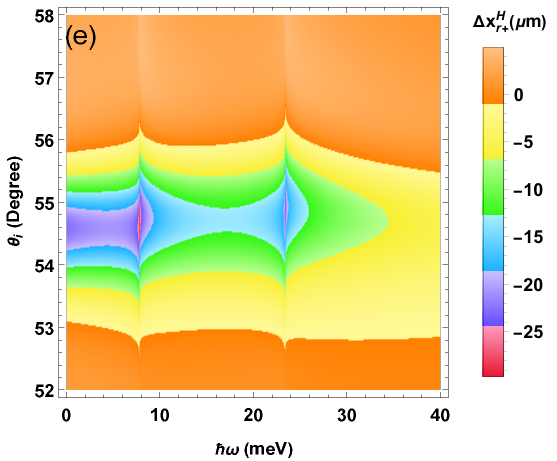}
\includegraphics[width=0.485\linewidth]{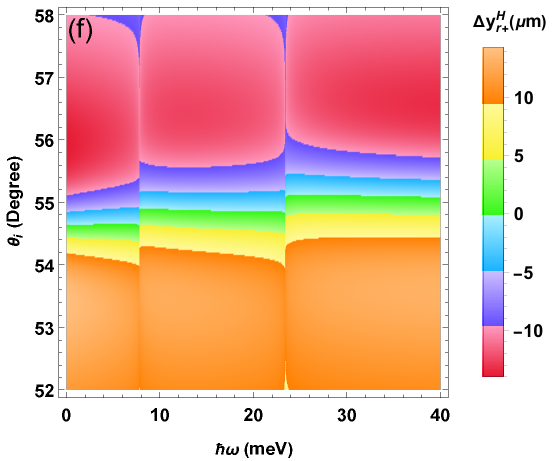}
\caption{Color maps of the $\Delta x_{r+}^{H}$ and  $\Delta y_{r+}^{H}$ versus $\hbar\omega$ and $\theta_{i}$  for the spin-up and spin-down electrons at the $K$ valley in distinct topological phases. (a) $\Delta x_{r+}^{H}$ and (b) $\Delta y_{r+}^{H}$ in the QSHI phase. (c) $\Delta x_{r+}^{H}$ and (d) $\Delta y_{r+}^{H}$ in the VSPM instant. (e) $\Delta x_{r+}^{H}$ and (f) $\Delta y_{r+}^{H}$ in the BI phase. We use $t_2 = \Delta_{so}/3\sqrt{3}$, $\Delta_{so}=3.9$ meV, $\mu_{F}$=0 eV,  $\mathcal{M}=\Delta_{z}$ and $\phi=+\pi/2$ ($\phi=-\pi/2$) for the spin-up (spin-down). }
\label{fig5}
\end{figure}

To probe the topological phase transitions in the PSHE due to the interaction of a Gaussian beam with the buckled Xene monolayer materials, Figs.~\ref{fig5}(a)-(f) depict the color maps of the LHCP longitudinal and transverse shifts versus photon energy and incident angles in the three distinct phases, QSHI, VSPM, and BI. Figs.~\ref{fig5}(a) and (b) demonstrate the PSHE of the buckled Xene monolayer material in the QSHI phase, illustrating selective excitation of spin-up and spin-down electrons in the $K$ valley.  The spin-valley coupled longitudinal spatial shift $\Delta x_{r+}^{H}$ exhibits extreme values at $\hbar\omega=\Delta_{so}$ and $\hbar\omega=3\Delta_{so}$ in the vicinity of Brewster’s
angles, $\theta_{B}$=$54.6^\circ$. At $\hbar\omega=\Delta_{so}$, $\Delta x_{r+}^{H}$ is positive and becomes negative at second excitation photonic energy $\hbar\omega=3\Delta_{so}$ because the real part of the Hall conductivity switch sign from positive to negative. Figure~\ref{fig5}(b) depicts the behavior of the spin and valley polarized transverse shifts $\Delta y_{r+}^{H}$ versus the incidence angle and photon energy in the QSHI phase. We observe positive and negative spatial spin-dependent shifts at the optical transition frequencies in the proximity of Brewster’s angle. The $\Delta y_{r+}^{H}$ is positive (negative) when $\theta_i < \theta_B$ ($\theta_i > \theta_B$). The jump-like behavior is shown on the first and second optical transitions.  The color maps of the spin-valley coupled LHCP $\Delta x_{r+}^{H}$ and $\Delta y_{r+}^{H}$ in the VSPM state are shown in Figs.~\ref{fig5}(c) and (d). In this phase, as mentioned above, we have only one optical transition at $\hbar\omega=4\Delta_{so}$. Therefore, we can observe negative $\Delta x_{r+}^{H}$ shift and step-like reflected
$\Delta y_{r+}^{H}$ at the optical transition energy. Figures~\ref{fig5}(e)  and (f), display the color maps of the longitudinal and transverse displacements in the BI phase, respectively.  We observe significantly
large negative shifts in this phase. The giant shifts can be seen at $\hbar\omega=\Delta_{so}$ and $\hbar\omega=6\Delta_{so}$ near the Brewster's angle, respectively.  In this phase, the width of the steps in $\Delta y_{r+}^{H}$ increases compared to the QSHI phase as the conductivity jumps originated by the spin-up electron are red-shifted, and the spin-down electron are blue-shifted.

\subsection{Transition metal dichalcogenide monolayers}
\begin{figure}[]
	\centering		
 	\includegraphics[width=0.4850\linewidth]{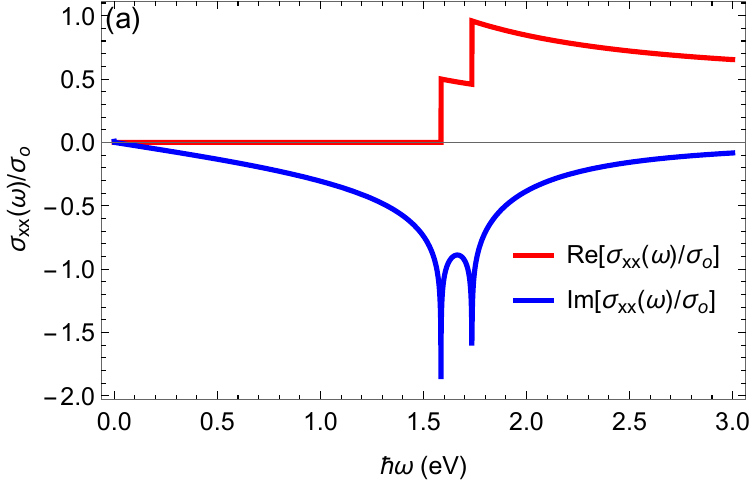}
 \includegraphics[width=0.4850\linewidth]{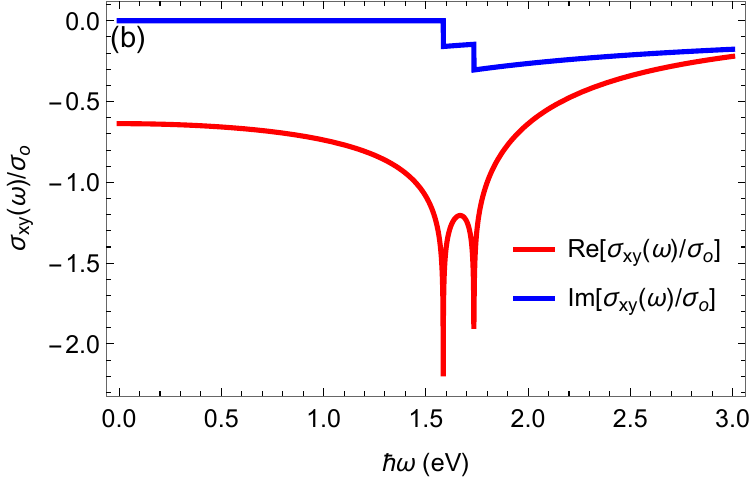}\\
 	\includegraphics[width=0.4850\linewidth]{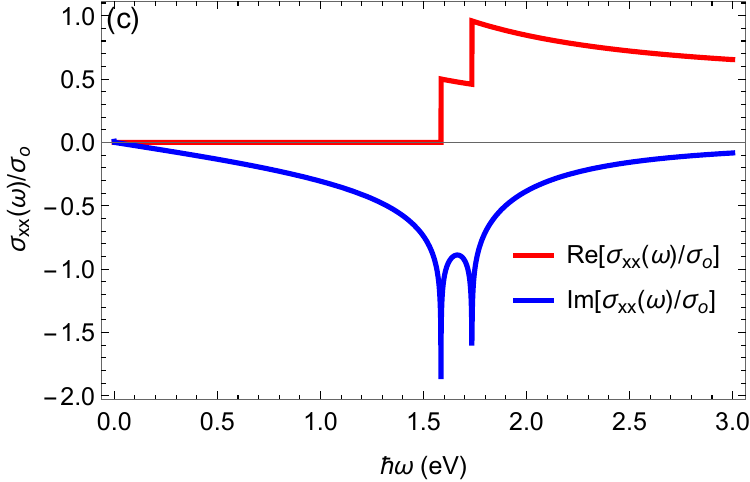}
 \includegraphics[width=0.4850\linewidth]{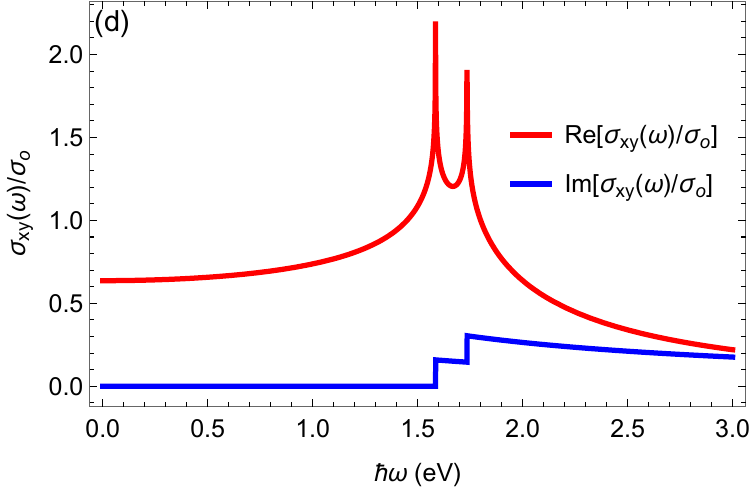}
	\caption{(a) The real and imaginary parts of $\sigma_{xx}(\omega)/ \sigma_{o}$ and (b) $\sigma_{xy}(\omega)/ \sigma_{o}$ as a function of photon energy for the $K$ valley. (c) Real and imaginary parts of $\sigma_{xx}(\omega)/ \sigma_{o}$ and (d) $\sigma_{xy}(\omega)/ \sigma_{o}$ as a functions of photon energy for the $K'$ valley.  For these simulations, we use $t_2 = \Delta_{\mathrm{TMD}}/3\sqrt{3}$, $\Delta_{\mathrm{TMD}}=75$ meV, $\mu_{F}$=0 eV, $\Delta$=1.66 eV,  $\mathcal{M}=\Delta/2$ and $\phi=+5\pi/6$ for spin-up and $\phi=-\pi/6$ for spin-down.}
	\label{fig6}
\end{figure}

Monolayer and few-layer transition metal dichalcogenides
((e.g. $\mathrm{MX_2}$, $\mathrm{M = Mo,~W;~ X = S,~Se}$)) have attracted huge attention for spintronic,
valleytronic, and optoelectronic applications \cite{geim2013van,xu2014spin,mak2016photonics}. Clearly,
the monolayer $\mathrm{MX_2}$ has a larger band gap in the near-infrared to the visible region.
$\mathrm{MX_2}$ materials are of particular interest
because they have a valley degree of freedom. Additionally, it has opposite spins at the two
in-equivalent $K$ and $K'$ valleys. By adjusting $\mathcal{M}=\Delta/2$, 
$t_2 = \Delta_{\mathrm{TMD}}/3\sqrt{3}$,  and $\phi=+5\pi/6 (-\pi/6)$, we can reproduce the dispersion relation of  monolayer $\mathrm{MX_2}$ materials at the $K$ and $K'$ valleys for both spins as 
$\mathcal{E}^{\tau, s}_{\eta}(\boldsymbol{k})=
 \tau s \Delta_{\mathrm{TMD}}/2+\eta \sqrt{\left(\hbar v_F \boldsymbol{k}\right)^2+(\Delta_{\tau}^{s})^2}$
where, $\Delta_{\tau}^{s}=0.5\Delta-0.5\tau s \Delta_{\mathrm{TMD}}$ is the energy gap in the monolayer $\mathrm{MX_2}$ materials. 
The splitting of the spin-up and spin-down
electrons at the valence bands occurs in each valley. At the $K$ valley, the upper (lower) valence band is occupied by spin-up (spin-down) electrons. Interestingly, due to the time-reversal symmetry, the spin splitting has opposite signs at $K$ and $K'$ valleys. Figures \ref{fig6}(a) and (b) present the real and imaginary parts of ${\sigma}_{xx}$  and ${\sigma}_{xy}$ of monolayer $\mathrm{MoS_{2}}$, for spin-up and spin-down electrons in the $K$ valleys, respectively. As shown, two jumps can be seen in the conductivity spectra at the optical excitation frequencies of the spin-up and spin-down electrons of the $K$ valley \cite{PhysRevB.89.115413}. Similarly, the optical conductivities ${\sigma}_{xx}$  and ${\sigma}_{xy}$ of  monolayer $\mathrm{MoS_{2}}$ in the $K'$ valley in Figs.~\ref{fig6}(c) and (d). 

\begin{figure}[]
	\centering		
	\includegraphics[width=0.485\linewidth]{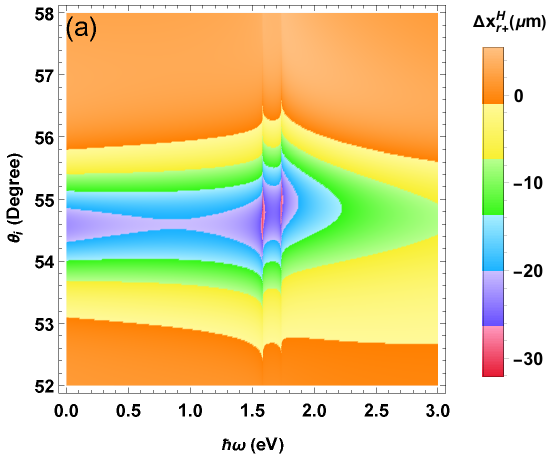}
	\includegraphics[width=0.485\linewidth]{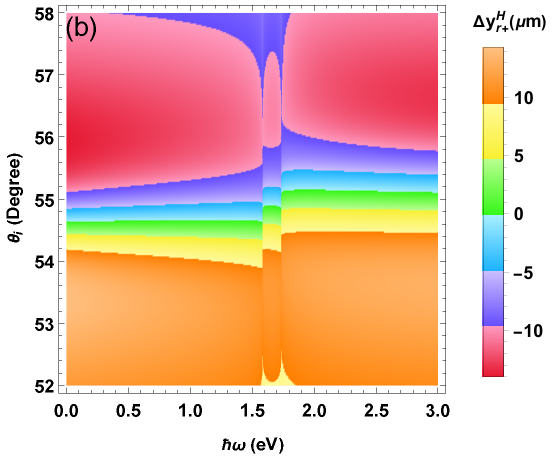}\\
	\includegraphics[width=0.485\linewidth]{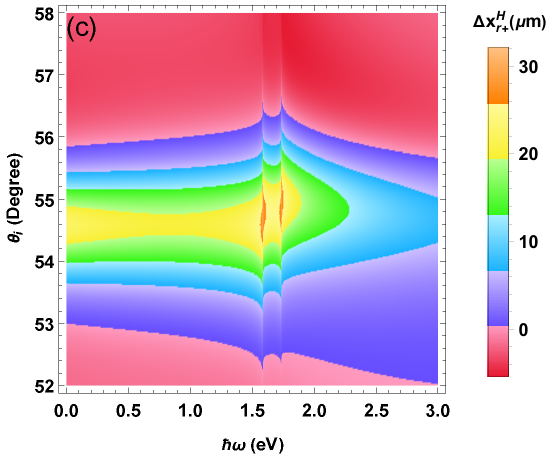}
	\includegraphics[width=0.485\linewidth]{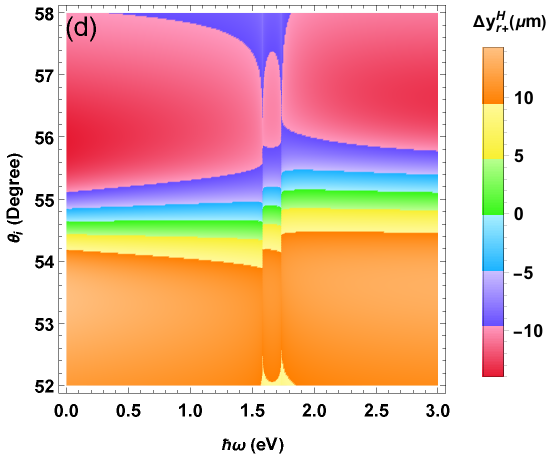}
\caption{Color maps of the $\Delta x_{r+}^{H}$ and  $\Delta y_{r+}^{H}$ versus $\hbar\omega$ and $\theta_{i}$  for the spin-up and spin-down electrons. (a) $\Delta x_{r+}^{H}$ and (b) $\Delta y_{r+}^{H}$ in the $K$ valley. (c) $\Delta x_{r+}^{H}$ and (d) $\Delta y_{r+}^{H}$ in the $K'$ valley. }
\label{fig7}
\end{figure}

Figures~\ref{fig7}(a) and (b) give the color maps of the spin and valley polarized $\Delta x_{r+}^{H}$ and $\Delta y_{r+}^{H}$ shifts for the $H$ polarized wave versus energy and incident angle for the $K$ valley.  Previously,  the reflected spin-dependent displacement in the air-glass interface has been experimentally demonstrated
to happen around the Brewster angle of $56.3^\circ$, with the enhancement of the original displacement by nearly four orders of magnitude \cite{hosten2008observation}. From Figs.~\ref{fig7}(a) and (b), we can see that there are some similar features in
the monolayer $\mathrm{MoS_{2}}$ and the insulating silicene systems (see Figs.~\ref{fig5}(e) and (f)). The prominent
reflected spin-dependent displacements occur around $\theta_{B}$ as usual.  It is seen that $\Delta x_{r+}^{H}$ gives extreme values in the vicinity of $\theta_{B}$ at which the $r_{pp}$ exhibits local minimums.  The large negative in-plane shift $\Delta x_{r+}^{H}$ for the $K$ valley can be seen in Fig.~\ref{fig7}(a). The negative shift can be attributed to the negative Hall conductivity as shown in Fig.~\ref{fig6}(b). Figure~\ref{fig7}(b) show variations of $\Delta y_{r+}^{H}$ with respect to the $\theta_{i}$ and $\hbar\omega$. We find that a step-like reflected spin-dependent shift is observed in $\mathrm{MoS_{2}}$ in the $K$ valley for the spin-up and spin-down electrons. Such a step-like behavior is inherited from the $\sigma_{xx}$ and $\sigma_{xy}$
parts of the optical conductivity as shown Figs.~\ref{fig6}(a) and (b). The transverse spin-dependent shift is positive for $\theta_{i}<\theta_{B}$ and becomes negative for $\theta_{i}>\theta_{B}$. The reflected spin-dependent shifts in the $K'$ valley can be treated similarly.  Figures~\ref{fig7}(c) and (d) show color maps of the reflected spin-dependent shift spectra as a function of photon energy and incident angle for $H$ polarized wave in the $K'$ valley, respectively. Interestingly, for the $K'$ valley, the longitudinal shift $\Delta x_{r+}^{H}$ is positive in the vicinity of $\theta_{B}$. The sign switching of the longitudinal shifts is closely associated with the transverse Hall conductivity as shown in Figs.~\ref{fig6}(c) and (d). For the $K$ and $K'$ valleys, $\sigma_{xy}$ switch signs from negative to positive, respectively. Therefore, both $\sigma_{xx}$ and $\sigma_{xy}$ collectively determine the behaviors of the reflected spin-dependent shifts in the Haldane materials in different valleys. It is seen that large in-plane photonic spin Hall shifts appear in monolayer $\mathrm{MoS_{2}}$, when compared with other 2D Haldane materials.

\begin{table*}[t]\label{mytab2}
\caption{Comparison of the PSHE in various 2D materials. The PSHE shifts are presented in $\mu$m, with the photon wavelength in units of $\mu$m, the effective band mass normalized by the bare electron mass $m_0$, and the effective band gap in eV at the average electron density $n=2\times 10^{12}$ nm$^{-2}$. In gapped Dirac materials, we simplify the dispersion relation as $\mathcal{E}\approx \Delta+\hbar^2 k^2/2 m^*$ for small $k$ values to extract the effective band masses. \label{mytab2}}
	\centering
\begin{ruledtabular}
 \begin{tabular}{lllll}
  2D materials & \text{$\Delta x^{H/V}_{r+}  /~~\Delta y^{H/V}_{r+}$} $(\mu m) $& Wavelength $(\mu m)$ & Band Mass ($m_0$)& Band Gap (eV)\\
		\hline 
  Graphene \cite{wu2020actively} & 6$\lambda=9.3 $ & 1.550 &-&0  \\ 
  Silicine \cite{wu2020weak}& 2$\lambda=1.62 $ & 0.810 & 0.0033 &0.012\\
  Anisotropic 2D crystals \cite{Zhang:18} & 1.5$\lambda=18.6$ & 12.4  &$m_x$=0.2, $m_y$=0 & Large\\
  Twisted few-layer anisotropic 2D crystals \cite{zhang2022photonic} & 6$\lambda=18.6 $  & 3.1 & $m_x$=0.2, $m_y$=0 &Large\\
  Bilayer borophene metasurfaces \cite{cheng2022giant} & -300$\lambda=-292.8 $& 0.974  &$m_x$=1.4, $m_y$=3.4  &0\\  $\alpha$-$\mathrm{Li_{3}N}$-type semimetals \cite{jia2022spin} & 0.17$\lambda=0.105$ &0.62 &0.044 &1.04\\
  Topological (This work) & 3$\lambda=30 $ & 10  &0.182 &0.52\\
  Normal insulator (This work) & -5$\lambda=-50 $  &  10& 0.351& 1.0 \\
  Buckled Xene  QSHI (This work) & -2.5$\lambda=25 $  & 10 &0.0033 & 0.012 \\
  Buckled Xene  VSPM (This work) & -3$\lambda=-30 $  & 10  & 0.0045&0.0156\\
  Buckled Xene  BI (This work) & -3$\lambda=-30 $ &  10 & 0.0068&0.0234\\
   2D  $\mathrm{MoS_{2}}$ (This work) & $\pm3.5\lambda=\pm 35 $  &  10 &0.53 &1.74\\
 \end{tabular}
 \end{ruledtabular}
\end{table*}
\section{CONCLUSION}\label{section:5}
The current study systematically discusses the photonic spin Hall effect arising from the interaction of a linearly polarized Gaussian beam with diverse 2D hexagonal materials exhibiting broken $\mathcal{T}$ and $\mathcal{I}$ symmetries. Notably, we emphasize the strong dependence of the photonic spin Hall effect on the non-trivial and trivial topologies inherent in these materials. In the first scenario, we meticulously examined the longitudinal and transverse displacements within both topological and trivial systems, elucidating the sign-switching phenomenon in the photonic spin Hall effect. Subsequently, we extended our exploration to the buckled Xene monolayer materials, underscoring the spin-dependent shifts as intricate indicators of topological phase transitions, each manifesting distinct behaviors in various states. Finally, the spin and valley-polarized reflected spin-dependent displacements in monolayer transition metal dichalcogenides were addressed. Our findings highlight the sensitivity of the photonic spin Hall effect to spin and valley indices, as well as to the effective mass bands within these materials.

Since several works on the photonic spin Hall effect in 2D systems are available, a proper comparison with those results seems to be in order (see Table \ref{mytab2}). Zhou, \emph{et al.} \cite{zhou2018photonic} have found that the spin-dependent splitting in the photonic spin Hall effect in graphene sheets is sensitive to the refractive index change of the sensing medium by considering the surface plasmon resonance effect and found that the maximum absolute of $\Delta x_{r+}$ is about $140\,\lambda$. Zhang \emph{et al.} \cite{Zhang:18} have studied the photonic spin Hall effect on the surface of anisotropic 2D atomic crystals with a parabolic dispersion relation. The band masses along the $x$ and $y$ directions were $0.2\,m_0$ and $m_0$, respectively, where $m_0$ is the free-electron mass. The peak of the photonic spin Hall effect was around $1.5\,\lambda$. Cheng \emph{et al.} \cite{cheng2022giant} studied theoretically PHSE shifts in bilayer borophene by making use of semiclassical optical conductivity shown a giant PSHE of the transmitted beams around $300\,\lambda$. In their study, they used the electronic relaxation time to be 65 fs, the band masses are $m_x=1.4\,m_0$, and $m_y=3.4\,m_0$ where $m_0$ denotes the static electron mass. The band gap in the system is about 1 eV. Jia \emph{et al.} \cite{jia2022spin} have investigated the photonic spin Hall effect of the transmitted light through thin films of $\mathrm{\alpha-Li_{3}N}$-type topological semimetals and found the maximum peak of the photonic spin Hall effect around $2\lambda$. In their system, the band gap is $1.04$ and the band mass is $0.044\, m_0$.

As summarized in Table \ref{mytab2}, our results indicate a significant increase in the photonic spin Hall effect $\Delta x^{H/V}_{r+}$ or $\Delta y^{H/V}_{r+}$ around 30 $\mu$m at given average electron densities. 
These comparisons indicate that our results align with previous work where applicable. Moreover, our results also extend to several relevant cases that were not previously discussed, for example, the spin and valley-dependent PSHE in TMDCs. 
We have shown that the magnitude of the photonic spin Hall effect, apart from the spin and valley indices, depends on material parameters namely, the effective band masses in the conduction band, the Fermi velocity as well as the topology of the band structure. Furthermore, the maximum photonic spin Hall effect in a 2D system seemingly occurs when the effective band masses are relatively large. Moreover, we propose that employing weak measurement approaches facilitates the experimental probing of the valley and spin-polarized Hall conductivity via the photonic spin Hall effect in Haldane materials. This opens up avenues for further exploration and application of these intriguing phenomena in the realm of optical and electronic studies.

\section{acknowledgments}
We acknowledge the financial support from the postdoctoral research grant YS304023905 and the NSFC under Grants No. 12174346.  R. A. received partial funding from the Iran National Science Foundation (INSF) under project No. 4026871.

\renewcommand{\bibname}{References}

\bibliography{References}
\end{document}